\documentclass[%
 reprint,
superscriptaddress,
nofootinbib,
 amsmath,amssymb,
 aps,
]{revtex4-2}

\usepackage[dvipsnames]{xcolor}
\usepackage{graphicx}
\usepackage{dcolumn}
\usepackage{bm}

\definecolor{brown}{RGB}{200,150, 50}
\definecolor{bgreen}{RGB}{0,150, 100}
\usepackage[dvipsnames]{xcolor}

\date{Accepted XXX. Received YYY; in original form ZZZ}

\begin{document}

\preprint{APS/123-QED}

\title{Temperature and Strong Magnetic Field Effects in Dense Matter}

\author{J. Peterson}
\affiliation{Department of Physics, Kent State University, Kent, OH 44243, USA}
\author{P. Costa}
\affiliation{CFisUC, Department of Physics, University of Coimbra, P-3004 - 516 Coimbra, Portugal}
\author{R. Kumar}
\affiliation{Department of Physics, Kent State University, Kent, OH 44243, USA}
\author{V. Dexheimer}
\affiliation{Department of Physics, Kent State University, Kent, OH 44243, USA}
\author{R. Negreiros}
\affiliation{Instituto de Fisica, Universidade Federal Fluminense, Niteroi, Brazil}
\author{C. Providência}
\affiliation{CFisUC, Department of Physics, University of Coimbra, P-3004 - 516 Coimbra, Portugal}

\date{\today}

\begin{abstract}
{We study consistently the effects of magnetic field on hot and dense matter. In particular, we look for differences that arise due to assumptions that reproduce the conditions produced in particle collisions or astrophysical scenarios, such as in the core of fully evolved neutron stars (beyond the protoneutron star stage). We assume the magnetic field to be either constant or follow a profile extracted from general relativity calculations of magnetars and make use of two realistic models that can consistently describe chiral symmetry restoration and deconfinement to quark matter, the Chiral Mean Field (CMF) and the Polyakov-loop extended Nambu-Jona-Lasinio (PNJL) models. We find that net isospin, net strangeness, and weak chemical equilibrium with leptons can considerably change the effects of temperature and magnetic fields on particle content and deconfinement in dense matter. We finish by discussing the possibility of experimentally detecting quark deconfinement in dense and/or hot matter and the possible role played by magnetic fields.}
\end{abstract}



\maketitle

\section{Introduction}

In the past decades, much research has been dedicated to dense and hot matter in the context of both particle colliders and astrophysics. The effects of strong magnetic fields {have} also been explored, but usually either at zero/small baryon chemical potential (or density) or {(effectively)} zero temperature. {This is because the heavy-ion collisions that create strong magnetic fields (necessary to significantly affect strongly interacting matter) require so much energy, that in this case the quarks in the Lorentz-contracted nuclei, which are moving practically at the speed of light, undergo only very weak forward scattering. The energy deposited behind them creates a ``fireball'', which is initially gluon dominated and evolves into a quark-gluon plasma which has nearly zero net baryon density (same amount of particles and anti-particles) (see Refs.~\cite{Braun-Munzinger:2015hba,Busza:2018rrf} for reviews).} These experiments take place at RHIC and LHC and can produce magnetic fields of the order of $m_\pi^2{/e}$, which translates to {$\sim$ $3\times10^{18}$} G \footnote{{Using Gaussian natural units, where the $\sqrt{4\pi}$ appears in the energy-momentum tensor, $1$ $\rm{MeV}^2 = 1.44\times 10^{13}$ G. Using Lorentz-Heaviside units, where the $\sqrt{4\pi}$ does not appear in the energy-momentum tensor, $1$ $\rm{MeV}^2 = 5.11\times 10^{13}$ G.}} or higher \cite{Deng:2012pc,Taghavi:2013ena,Tuchin:2013apa}. 

In neutron stars, {the ratio of temperature to Fermi energy is incredibly small ($\sim0.001\%$) justifying the approximation $T\sim0$.} Soft gamma
repeaters (SGRs) and anomalous X-ray pulsars (AXPs), both named according to their unusual electromagnetic emission characteristics, 
present the strongest magnetic fields inferred at the stellar
surface, reaching $10^{15}$ G \cite{Kaspi:2017fwg}. Complementary, data from the source 4U 0142+61 for slow phase modulations in
hard X-ray pulsations
suggests magnetic fields of the order of $10^{16}$ G { \cite{Makishima:2014dua}} inside this pulsar. Because the maximal magnetic field in the interior of neutron stars cannot be measured directly, it is estimated using the virial theorem as a theoretical upper limit, providing strengths of the order of $10^{18}$ G \cite{1991ApJ...383..745L}.

At finite, but not extremely large temperature ($0<T<100$ MeV), strong magnetic fields in dense matter have not been explored in detail. The main reason being that until recently, there was no physical system that had been detected with properties corresponding to {those} {conditions}. This changed in 2017, when the first gravitational waves from a neutron star merger were measured by LIGO/Virgo \cite{LIGOScientific:2017vwq}. After that, dense and hot environments in which the magnetic field can be strong became much more tangible. Several works on mergers of typical neutron stars (without strong magnetic fields) can produce magnetic fields of the order of $10^{16}$ G {\cite{Giacomazzo:2014qba,Dionysopoulou:2015tda,Harutyunyan:2018mpe,Ciolfi:2019fie,Most:2019kfe,Ciolfi:2020cpf,Palenzuela:2021gdo,Ruiz:2021qmm,Suvorov:2021ymy}}, with higher values expected from mergers of neutron stars with strong magnetic fields, which {have not yet been simulated}.

Ignoring temperature and magnetic fields at first (for simplicity), the core of neutron stars can reach densities of several times nuclear saturation density. In this regime, simple back of the envelope calculations show that hadrons (protons, neutrons, hyperons) are already overlapping, and a description that takes the inner composition of hadrons into account has to be employed. When temperature is included, taking deconfinement to quark matter into account becomes much more important, as the chemical potential at which deconfinement takes place is expected to be lower (see Fig.~1 of Ref.~\cite{Alford:2007xm} for a typical shape of the Quantum chromodynamics, QCD, phase diagram). {Perfect examples of such conditions are neutron-star mergers, that not only are extremely dense (surpassing the chemical potentials of the inspiriling stars that merged \cite{Most:2019onn}), but also hot. For this reason, it is crucial that temperature is included microscopically in EoS models, allowing the degrees of freedom to change as a function of temperature.} Note that, even within the hadronic phase, it has been shown that for realistic EoS's thermal effects cannot be reproduced with simple approaches, e.g. use of a constant thermal or adiabatic index law \cite{Raduta:2022elz,Kochankovski:2022ygd}. 

With all of these different systems (and respective conditions) in mind, we study in detail in this work dense and hot matter under strong magnetic fields. We make use of two relativistic models (ensuring a causal behavior \footnote{{Relativistic} models are causal as long as the vector interactions are not too strong, which is the case in this work.}) that include deconfinement to quark matter. The first one is the {Chiral Mean Field (CMF) model \cite{Papazoglou:1998vr}} and the second one is the {Polyakov-loop extended Nambu–Jona-Lasinio (PNJL) model \cite{Meisinger:1995ih,Fukushima:2003fw}}, both of which will be described in the next section, after {discussing} different conditions related to conservation laws and different kinds of equilibria. Both of these models are also chiral, {in which case the masses of baryons and quarks are not ``bare'' masses modified by the medium, but instead are fully (or almost fully) generated from interactions with the medium. These ``effective'' masses decrease from vacuum values (for baryons) or constituent values (for quarks) to smaller values as a function of density and/or temperature.}

{Up to this point, few works utilize an EoS for dense matter simultaneously accounting for magnetic field and temperature effects. Two approaches dominate the literature in the field. The first is to approximate the system as a relativistic Fermi gas of baryons and leptons using Walecka-type models or simply using a relativistic free Fermi gas  without strong force interactions \cite{Strickland:2012vu, Ferrer:2019xlr, Rabhi_2011, Isayev:2011zz}. The other is using the NJL/PNJL or MIT bag model to study quark systems and their properties in neutron-star or heavy-ion collision conditions \cite{Dexheimer:2012mk, fune2019magnetized, Gyory:2022hnv, Mishra:2022pee, Abhishek:2018xml, Chatterjee:2011ry, Gorbar:2011ya, Frolov:2010wn, Andersen:2012bq}, and to examine the QCD phase diagram \cite{Mizher:2010zb,Avancini:2011zz, Costa:2015bza, Ferreira:2013tba, Ferreira:2014kpa, Ferreira:2017wtx}. Additionally, the Thomas-Fermi model has been used to study atoms as a Fermi gas \cite{Thorolfsson:1997dq, Lai_2001} (this has implications for neutron star crusts), a Walecka-type model has been used to study the neutron star crust-core transition \cite{Ferreira:2021pam}, the multiple reflection expansion framework has been used to investigate the surface tension of quark matter droplets with neutron-star conditions \cite{Lugones:2018qgu}, and macroscopic properties of magnetars have been studied using a current density influenced by both temperature and magnetic field \cite{Turolla_2015}. Let us mention that transport coefficients of  hot and dense hadronic matter \cite{Das:2019pqd} and quark matter \cite{Fukushima:2017lvb} in the presence of magnetic field have also been studied.}

\section{Formalism}\label{Formalism}

\subsection{Different Conditions}
To describe different systems, from laboratory experiments to astrophysical objects, we start by defining some relevant conditions:
\begin{itemize}
\item isospin symmetry: due to the extremely short duration of heavy-ion experiments {($\sim10$ fm/c $\sim10^{23}$ s)}, there is not enough time to create net isospin {through weak reactions}, and the isospin or charge fraction remains that of the initial nuclei (conservation of isospin). {For the case of extremely high-energy collisions, when the nuclei pass straight through one other, matter produced in the fireball has no net isospin (this is also the simplest case to describe and the ``canonical'' one for heavy-ion collisions)}
\begin{equation}
Y_I=\frac{I}{{\mathfrak{B}}}\sim 0\ \ \rm{or}\ \ Y_Q=\frac{Q}{{\mathfrak{B}}}\sim 0.5 ,
\end{equation}
where $I$ is the total isospin, $Q$ the hadron/quark electric charge, and {$\mathfrak{B}$} the number of hadrons and quarks (note that quarks have baryon number $1/3$). {In this work, this} is achieved through the assumption of equilibrium with respect to isospin or charge
\begin{equation}
\mu_I=0\ \ \rm{or}\ \ \mu_Q=0,
\end{equation}
where $\mu$ is the chemical potential \footnote{The relation between Eqs. (1) and (2) is not straight forward in the presence of strong magnetic fields. See discussion in the end of Section III A.}. The conservation of isospin and electric charge fractions are equivalent, as long as the strangeness is zero \cite{Aryal:2020ocm} (see next item);
\item zero net strangeness: due to the extremely short duration of heavy-ion experiments, there is also not enough time to create net strangeness {through weak reactions}, and the {net} strangeness fraction remains that of the initial nuclei (conservation of strangeness)
\begin{equation}
Y_S=\frac{S}{{\mathfrak{B}}}=0 ,
\end{equation}
where $S$ is the total {net} strangeness. This is achieved by introducing an independent chemical potential $\mu_S$;
\item charge neutrality (with leptons): astrophysical objects are {understood to be electrically charge neutral, as the electromagnetic force is much stronger than gravity}. This is numerically enforced in models by ensuring that leptons, typically electrons and muons, balance the charge of hadrons, typically protons, neutrons, hyperons, and quarks (conservation of electric charge)
\begin{equation}
Y_{\rm{lep}}=Y_Q .
\end{equation}
\item chemical equilibrium with leptons: in fully evolved neutron stars, {beyond the protoneutron star stage,} weak chemical equilibrium is reached with the leptons. The neutrinos escape, $\mu_\nu=0$, and  the chemical potential of hadrons/quarks and leptons relate through
\begin{equation}
\mu_e=\mu_\mu=-\mu_Q ,
\end{equation}
where
\begin{equation}
\mu_Q=\mu_p-\mu_n\ \ \rm{or}\ \ \mu_Q=\mu_u-\mu_d ,
\end{equation}
(see Appendix A of Ref.~\cite{Aryal:2020ocm} for a full list of chemical potential relations);
\item chemical equilibrium with respect to strangeness: in fully evolved neutron stars, weak chemical equilibrium is also achieved with respect to strangeness
\begin{equation}
\mu_S=0 .
\end{equation}
\end{itemize}

To describe magnetic fields, we assume two possibilities:
\begin{itemize}
\item constant magnetic field: due to the very small {size and} time scale of heavy-ion collisions, anisotropies in the magnetic field are not relevant for our {exploratory} discussion and, therefore, can be disregarded
\item magnetic field profile: inside stars of $\sim12$ km radius, spacial magnetic field anisotropies are extremely relevant. Therefore, we assume a magnetic field profile as a function of baryon chemical potential $\mu_B$ and dipole magnetic moment $\mu$ extracted from realistic general relativity calculations that also fulfill Maxwell equations (including conservation of electric charge and magnetic flux) \cite{Dexheimer:2016yqu}
\begin{eqnarray}
B^*(\mu_B)=\frac{(a + b \mu_B + c \mu_B^2)}{B_c^2} \ \mu,
\label{3}
\end{eqnarray}
with coefficients $a=-7.69\times10^{-1}$ $\frac{\rm{G}^2}{\rm{A m}^2}$, $b=1.20\times10^{-3}$ $\frac{\rm{G}^2}{\rm{A m}^2 \rm{MeV}}$, and $c=-3.46\times10^{-7}$ $\frac{\rm{G}^2}{\rm{A m}^2 \rm{MeV}^2}$. 

Eq.~(\ref{3}) requires $\mu_B$ in MeV and $\mu$ in Am$^2$ in order to produce $B^*$ in units of the critical field for the electron $B_c=4.414\times 10^{13}$ G. This profile corresponds to the magnetic field along the polar direction of a massive star, with different strength depending on the value chosen for $\mu$.
For this work, we choose four different magnetic field profiles, each generated from different values of $\mu$:
\begin{itemize}
\item[*] $\mu=3\times10^{32}$ A m$^2$;
\item[*] $\mu=6\times10^{32}$ A m$^2$;
\item[*] $\mu=12\times10^{32}$ A m$^2$;
\item[*] $\mu=24\times10^{32}$ A m$^2$,
\end{itemize}
which will henceforth be identified as ``profile 3," ``profile 6," ``profile 12," and ``profile 24," respectively.
\end{itemize}

\subsection{CMF Model}

In this subsection, we describe the SU(3) Chiral Mean Field (CMF) model. Spontaneous chiral symmetry breaking is related to the formation of scalar condensates ({typically}, isoscalar $\sigma$, isovector $\delta$, and isoscalar with hidden strangeness $\zeta$), which can be used as order parameters for symmetry breaking. {In hadronic chiral models,} these condensates are associated with scalar mesons that mediate the attraction between baryons (nucleons and hyperons). The description of equivalent vector mesons (isoscalar $\omega$, isovector $\rho$, and isoscalar with hidden strangeness $\phi$) mediate the repulsion between hadrons. Only the mean values of the mesons are used in the CMF model, as the meson field fluctuations are expected to be small at high densities. We further make use of a non-linear realization of the sigma model, which allows a very good agreement with low-energy nuclear data, such as the vacuum masses of the hadrons and the pion and kaon decay constants \cite{Papazoglou:1998vr}. Additional explicit symmetry breaking gives masses to the {pseudo-scalar} mesons. To describe neutron stars, a free gas of leptons is also included {and standard astrophysical properties are reproduced \cite{Dexheimer:2008ax,Dexheimer:2009hi,Roark:2018boj,Dexheimer:2018dhb}}. 

\begin{table*}[t!]
\centering
\caption{Table of anomalous magnetic moment couplings $k_i$ for all the particles included in the CMF model (obtained from Ref.~\citep{Zyla:2020zbs}).}
\def\arraystretch{1.8}
\begin{tabular}{ccccccccccccc}
\hline
\hline
p & n & $\Lambda$ & $\Sigma^+$ & $\Sigma^0$ & $\Sigma^-$ & $\Xi^0$ & $\Xi^-$ & e & $\mu$ & u & d & s \\
\hline
\ \ \ $1.79$\ \ \ &\ \ \ $-1.91$\ \ \ &\ \ \ $-0.61$\ \ \ &\ \ \ $1.67$\ \ \ &\ \ \ $1.61$\ \ \ &\ \ \ $-0.38$\ \ \ &\ \ \ $-1.25$\ \ \ &\ \ \ $0.06$\ \ \ &\ \ \ $0.00116$\ \ \ &\ \ \ $0.001166$\ \ \ &\ \ \ 0\ \ \ &\ \ \ 0\ \ \ &\ \ \ 0\ \ \ \\
\hline
\hline
\end{tabular}
\label{kappatable}
\end{table*}

Inspired by unified approaches for the liquid-gas phase transition \cite{Oertel:2016bki}, a unified approach for quark deconfinement {was} implemented in the CMF model. Unified means that all degrees of freedom are always included a priori in the description of both phases, allowing for different kinds of phase transition between the phases \footnote{Note that an alternative version of the CMF model includes in addition the chiral partners of the baryons and gives the baryons a finite size \cite{Steinheimer:2011ea, Motornenko:2019arp}}. This is done by including up, down, and strange quarks to the CMF model in a way similar to the baryons, as shown in the Lagrangian density of the model
\begin{eqnarray}
\mathcal{L} = \mathcal{L}_{\rm{Kin}} + \mathcal{L}_{\rm{Int}} + \mathcal{L}_{\rm{Self}} + \mathcal{L}_{\rm{SB}} - U ,
\end{eqnarray}
where $\mathcal{L}_{\rm{Kin}}$ is the kinetic energy density of hadrons and quarks, $\mathcal{L}_{\rm{Int}}$ describes the interactions between baryons and quarks mediated by the mesons, $\mathcal{L}_{\rm{Self}}$ describes the self-interactions of the scalar and vector mesons, $\mathcal{L}_{\rm{SB}}$ the chiral symmetry breaking term, and $U$ the effective potential for the scalar field $\Phi$, as shown below
\begin{eqnarray}
\mathcal{L}_{\rm{Kin}} &=&\sum_i \bar{\psi_i} \left[i\gamma^\mu (\partial_\mu + i q_i A^{EM}_{\mu})\right ] \psi_i , \nonumber\\
\mathcal{L}_{\rm{Int}} &=& -\sum_i \bar{\psi_i} \big[\gamma_0 \big(g_{i \omega} \omega + g_{i \phi} \phi + g_{i \rho} \tau_3 \rho \big) \nonumber\\&-& \frac{1}{2}\kappa_i \sigma^{\mu\nu}F_{\mu\nu} + M_i^* \big] \psi_i , \nonumber\\
\mathcal{L}_{\rm{Self}} &=& \frac{1}{2} \big(m_\omega^2 \omega^2 + m_\rho^2 \rho^2 + m_\phi^2 \phi^2\big) \nonumber \\
&+& g_4 \left(\omega^4 + \frac{\phi^4}{4} + 3 \omega^2 \phi^2 + \frac{4 \omega^3 \phi}{\sqrt{2}} + \frac{2 \omega \phi^3}{\sqrt{2}}\right) \nonumber \\
&-& k_0 \big(\sigma^2 + \zeta^2 + \delta^2 \big) - k_1 \big(\sigma^2 + \zeta^2 + \delta^2 \big)^2 \nonumber \\
&-& k_2 \left(\frac{\sigma^4}{2}+\frac{\delta^4}{2} + 3 \sigma^2 \delta^2 + \zeta^4\right) - k_3 \big(\sigma^2 - \delta^2 \big) \zeta \nonumber 
\end{eqnarray}
\begin{eqnarray}
&-& k_4 \ \ln{\frac{ \big(\sigma^2 - \delta^2 \big) \zeta}{\sigma_0^2 \zeta_0}} ,\nonumber  \\ 
\mathcal{L}_{\rm{SB}} &=& -m_\pi^2 f_\pi \sigma - \left(\sqrt{2} m_k^ 2f_k - \frac{1}{\sqrt{2}} m_\pi^ 2 f_\pi\right) \zeta , \nonumber 
\end{eqnarray}
\begin{eqnarray}
U &=& \big(a_o T^4 + a_1 \mu_B^4 + a_2 T^2 \mu_B^2 \big) \Phi^2 \nonumber \\
&+& a_3 T_o^4 \ \ln{\big(1 - 6 \Phi^2 + 8 \Phi^3 -3 \Phi^4 \big)} .
\label{upol}
\end{eqnarray}
The index $i$ runs over the baryon octet  and the three light quarks. $q$ is the electric charge, $g$ the coupling constant, and $M^*$ the effective mass of particle $i$. $A^{EM}_\mu$ accounts for the interaction with the external magnetic field. Choosing the magnetic field to point locally in the z-direction and the vector potential to be $A_{EM}^\mu=(0, -B y, 0, 0)$, implies $\frac{1}{2}\kappa_i \sigma^{\mu\nu}F_{\mu\nu}= \kappa B S_3$, where $S_3=\Big(\begin{smallmatrix}
\sigma_3 & 0\\
0 & \sigma_3 
\end{smallmatrix}\Big)$ using the notation of the Pauli matrices {and the anomalous magnetic moment (AMM) $\kappa_i$ is $k_i$, the AMM coupling strength (see Table \ref{kappatable} for values, {noting that the AMM for quarks is not taken into account in this work}), multiplied by the magneton. {The magneton for baryons is the nuclear magneton and for leptons it is calculated as $e/2M_{i, \rm{vacuum}}$.}}

{The scalar coupling constants of the hadronic part of the model were fitted to reproduce vacuum masses of baryons, the pion and kaon decay constants, and reasonable values for the hyperon potentials ($U_\Lambda=-28.00$ MeV, $U_\Sigma=5$ MeV, $U_\Xi=-18$ MeV) at saturation. The vector coupling constants of the hadronic part of the model reproduce the following nuclear properties: saturation density $\rho_0=0.15$ fm$^{-3}$, binding energy per nucleon $B/A=-16$ MeV, compressibility $K=300$ MeV, and symmetry energy $E_{\rm{sym}}=30$ MeV with slope $L=88$ MeV. The predicted critical point for the nuclear liquid-gas phase transition of isospin symmetric matter lies at $T_c=16.4$ MeV, $\mu_{B,c}=910$ MeV. The values of the coupling constants can be found in Ref.~\cite{Roark:2018uls}. Only mean-field mesons, which provide the interaction for hadrons and quarks, are included in this work and their masses are fixed to their vacuum values.} 

Concerning the potential $U$, its pure temperature contribution is fitted to reproduce the results of the Polyakov loop in the PNJL approach \cite{Ratti:2006ka, Roessner:2006xn} at zero baryon chemical potential, while the chemical potential and mixed terms are motivated by symmetry and simplicity. The former one also contains the correct scale in the asymptotic zero-temperature case. The coupling constants of the quark sector are fitted to lattice data and to expectations from the phase diagram. The lattice data include (i) the location of the first-order phase transition and the pressure functional $P(T)$ at $\mu_B=0$ for pure gauge (the latter resulting from the PNJL model fitted to lattice) \cite{Ratti:2005jh, Roessner:2006xn} and (ii) the crossover pseudo-critical temperature and susceptibility $d\Phi/dT$ at vanishing chemical potential, together with the location of the ($T,\mu_B$) critical end-point for zero net-strangeness isospin-symmetric matter \cite{Fodor:2004nz}. The phase diagram expectations include a continuous first-order phase-transition line that starts at $T=167$ MeV temperature for zero-strangeness isospin-symmetric matter and terminates on the zero-temperature axis at four times the saturation density of chemically-equilibrated and charge-neutral matter.

The transition from hadrons to quarks as the density and temperature increase is done by means of $\Phi$, named in analogy with the Polyakov loop \cite{Fukushima:2003fw}, introduced in the effective mass of baryons and quarks. When $\Phi$ is near $1$, the effective mass of baryons 
\begin{equation}
M_{B}^* = g_{B \sigma} \sigma + g_{B \delta} \tau_3 \delta + g_{B \zeta} \zeta + M_{0_B} + g_{B \Phi} \Phi^2 ,
\end{equation}
becomes too large for them to be populated, while the effective masses of quarks
\begin{equation}
M_{q}^* = g_{q \sigma} \sigma + g_{q \delta} \tau_3 \delta + g_{q \zeta} \zeta + M_{0_q} + g_{q \Phi}(1 - \Phi) ,
\end{equation}
become low enough for them to become relevant \cite{Dexheimer:2009hi}, with small bare masses $M_0$. This {setup gives rise to} first-order phase transitions (at zero and small temperatures), as well as crossovers (at large temperatures), as predicted by lattice QCD calculations \cite{Aoki:2006we} in that regime. {To reproduce crossovers, as the temperature goes up, quarks slowly start to appear at lower $\mu_B$'s. This includes quarks dissolved in the hadronic phase (and vice-versa). Regardless, quarks never appear close to the nuclear liquid-gas phase transition (see Tab.~2 of Ref.~\cite{Roark:2018boj}).}

The CMF model has already been used to study the effects of strong magnetic fields at zero temperature in neutron stars \cite{Dexheimer:2011pz,Franzon:2015sya,Dexheimer:2016yqu,Dexheimer:2021sxs,Marquez:2022fzh}, but it is used here to study finite temperature dense matter (with effects of strong magnetic fields) for the first time and dense matter under different conditions (with effects of strong magnetic fields) for the first time. 
Equations describing the effects of magnetic field in a Free Fermi gas at finite temperature can be found in Ref.~\cite{Strickland:2012vu,Peterson:2021teb}. There (and here) AMM {couplings, which give rise to imbalances of particles with different spin projections due to the magnetic field,} are also included, for both charge neutral and charged hadrons. Magnetic effects are not included in the mesons and $\Phi$ in the CMF model, as they are sub leading. The interactions we use in this work (in addition to the meson self interactions) appear in modifications of the  masses and fermion energy spectra of free fermions, as discussed recently in Refs.~\cite{Dexheimer:2021sxs,Marquez:2022fzh}. As a result of quantization of the orbits of charged particles in the presence of the magnetic field, Landau levels are populated until the density of a given level for particle $i$, {$n_{i,\nu'}$} goes to zero.  At finite temperature, this is numerically done populating Landau levels until the density of the level {$\nu'$} represents only a small fraction of the density of all levels combined
\begin{align}
{n_{i,\nu'}}\le 10^{-5} \sum_{\nu=0 \rm{~or} ~1}^{\nu'}{n_{i, \nu}}.
\end{align}

\subsection{PNJL Model}

In this subsection, we describe the SU(3) Polyakov-loop extended Nambu–Jona-Lasinio (referred to as PNJL) model. Like the CMF model, it is based on spontaneous and explicit chiral symmetry breaking but, to describe the generation of mass of the quarks. In this case, the condensates are explicitly tied to each of the three quarks. Additionally, the quarks couple to a (spatially constant) temporal background gauge field, represented in terms of the Polyakov loop \cite{Meisinger:1995ih,Fukushima:2003fw,Ratti:2005jh}. The Lagrangian density is given by
\begin{eqnarray}
{\cal L} &=& {\bar{q}} \left[i\gamma_\mu D^{\mu}-
{\hat m}_f \right ] q ~+~ {\cal L}_{sym}~+~{\cal L}_{det} +\mathcal{U}\left(\Phi,\bar\Phi;T\right) ,\nonumber\\
\label{Pnjl}
\end{eqnarray}
where the quark sector is described by the  SU(3) Nambu--Jona-Lasinio model, which includes scalar-pseudoscalar and the t'Hooft six fermion interactions  \cite{Hatsuda:1994pi,Klevansky:1992qe}, with ${\cal L}_{sym}$ and ${\cal L}_{det}$ given by \cite{Buballa:2003qv}
\begin{eqnarray*}
	{\cal L}_{sym}&=& G \sum_{a=0}^8 \left [({\bar q} \lambda_ a q)^2 + 
	({\bar q} i\gamma_5 \lambda_a q)^2 \right ] ,\nonumber\\
	{\cal L}_{det}&=&-K\left\{{\rm det} \left [{\bar q}(1+\gamma_5) q \right] + 
	{\rm det}\left [{\bar q}(1-\gamma_5)q\right] \right \} ,
\end{eqnarray*}
where $q = (u,d,s)^T$ represents a quark field with three flavors, ${\hat m}_f= {\rm diag}_f (m_u^0,m_d^0,m_s^0)$ is the corresponding (current) mass matrix, $\lambda_0=\sqrt{2/3}I$,  where $I$ is the unit matrix in the three flavor space, and $0<\lambda_a\le 8$ denote the Gell-Mann matrices.
The coupling between the magnetic field $B$ and quarks, and between the effective gluon field and quarks is implemented  {\it via} the covariant derivative $D^{\mu}=\partial^\mu - i q_f A_{EM}^{\mu}-i A^\mu$, where $q_f$ represents the quark electric charge,  $A^{EM}_\mu$ accounts for the interaction with the magnetic field, and $A^\mu(x) = g_{strong} {\cal A}^\mu_a(x)\frac{\lambda_a}{2}$ where {$g_{strong}$ is the strong coupling} and ${\cal A}^\mu_a$ is the SU$_c(3)$ gauge field.
Considering once more  a magnetic field locally pointing in the $z$ direction, the vector potential is $A_{EM}^\mu=(0, -B y, 0, 0)$.

The trace of the Polyakov line defined by
$ \Phi = \frac 1 {N_c} {\langle\langle \mathcal{P}\exp i\int_{0}^{\beta}d\tau\,
A_4\left(\vec{x},\tau\right)\ \rangle\rangle}_\beta$
is the Polyakov loop, which is the {\it exact} order parameter of the $Z_3$ symmetric/broken phase transition in pure gauge. {$N_c$ is the number of colors, $A_4 = i A_0$ is the temporal component of the Euclidean gauge field $(\vec{A}, A_4)$, $\mathcal{P}$ denotes path ordering, and the usual notation $\beta = 1/T$ has been introduced.} In the presence of quarks, it becomes an approximate order parameter for quark deconfinement. To describe the pure gauge sector, the effective potential, $\mathcal{U}$ is chosen to reproduce the results obtained in lattice calculations \cite{Roessner:2006xn}
\begin{eqnarray}
\frac{\mathcal{U}}{T^4}&=& -\frac{a\left(T\right)}{2}\bar\Phi \Phi \nonumber\\
&+&\, b(T)\mbox{ln}\left[1-6\bar\Phi \Phi+4(\bar\Phi^3+ \Phi^3)-3(\bar\Phi \Phi)^2\right],\nonumber\\
\label{Ueff}
\end{eqnarray}
where $a\left(T\right)=a_0+a_1\left(\frac{T_0}{T}\right)+a_2\left(\frac{T_0}{T}\right)^2$ and $b(T)=b_3\left(\frac{T_0}{T}\right)^3$.
The standard choice of the parameters for the effective potential $\mathcal{U}$ is $a_0 = 3.51$, $a_1 = -2.47$, $a_2 = 15.2$, and $b_3 = -1.75$. The parameter $T_0$ is the critical temperature for the deconfinement phase transition within a pure gauge approach. It is fixed to a constant $T_0=270$ MeV, according to lattice findings.

The model being an effective one (up to the scale $\Lambda_{QCD}$) and not renormalizable, we use as a regularization scheme, a sharp cut-off, $\Lambda$, in 3-momentum space, only for the divergent ultra-violet integrals. 
The parameters of the model, $\Lambda$, the coupling constants $G$ and $K$, and the current quark mass for the strange quark $m_s^0$ are determined  by fitting {the decay constants and masses} $f_\pi$, $m_\pi$, $m_K$, and $m_{\eta'}$ to their experimental values in vacuum, while $m_u^0=m_d^0$ is fixed at 5.5 MeV. 
We consider then $\Lambda = 602.3 \, {\rm MeV}$, $m_u^0= m_d^0=\,  5.5 \,{\rm MeV}$, $m_s^0=\,  140.7\, {\rm MeV}$, $G \Lambda^2= 1.385$, and $K \Lambda^5=12.36$, as in {Ref.~}\cite{Rehberg:1995kh}.

In the mean field approximation the effective quarks masses are given by the gap equations
\begin{align}
\label{GapEqs}
\left\{
\begin{array}[c]{c}%
M_u=m_u-G \langle \overline{q}_u q_u\rangle-K \langle \overline{q}_d q_d\rangle\langle \overline{q}_sq_s\rangle ,\vspace{.1cm}
\\
M_d=m_d-G \langle \overline{q}_dq_d\rangle-K \langle \overline{q}_uq_u\rangle\langle \overline{q}_sq_s\rangle ,\vspace{.1cm}
\\
M_s=m_s-G \langle \overline{q}_sq_s\rangle-K \langle \overline{q}_uq_u\rangle\langle \overline{q}_dq_d\rangle ,
\end{array}
\right.
\end{align}
where the condensates are given by the following momentum integral
\begin{align}
\label{eqcondensate}
\langle\overline{q}_fq_f\rangle=-4 M_f \int\frac{\mathrm{d}^4p}{\left(2\pi\right)^4}\frac{1}{p_4^2+p^2+M_f^2}\,.
\end{align}

The extension to take into account the medium effects of finite temperature and/or chemical potential can be done by replacing the $p_4$ integration by a summation over Matsubara frequencies
\begin{align}
p_4 &\rightarrow \pi  T (2 n + 1) - i \mu , \nonumber\\
\int \mathrm{d}p_4 &\rightarrow  2\pi T \sum _{n=-\infty }^{+\infty }\,.
\end{align}
The effect of a finite magnetic field can then be seen as the substitution of the integration over transverse momentum, with respect to the local direction of the magnetic field, by a summation over Landau levels  ({related to} the index $m$) averaged over the spin  related index, $s$,
\begin{align}
  \int\frac{\mathrm{d}^2p_\perp}{\left(2\pi\right)^2}&
  \rightarrow \frac{2\pi\left|q\right|B}{\left(2\pi\right)^2} \frac{1}{2}\sum_{s=-1,+1}\sum_{m=0}^{+\infty},\nonumber\\
  \qquad p^2_\perp &
  \rightarrow (2 m+1-s)\left|q\right| B.
\end{align}

\section{Results}

In this section we focus our analysis on two kinds of matter combining the discussion from Section II A
\begin{itemize}
\item {\bf{neutron-star matter}}: charge neutral, in {weak} chemical equilibrium with leptons and with respect to strangeness. We investigate the effects of constant magnetic field and a {more} realistic magnetic-field profile;
\item {\bf{heavy-ion collision matter}}: isospin symmetric, with zero net strangeness. We investigate the effects of constant magnetic field.
\end{itemize}

\subsection{CMF Model}

\begin{figure}[t!]
\centering
\vspace{-.55cm}
\includegraphics[width=1.1\linewidth]{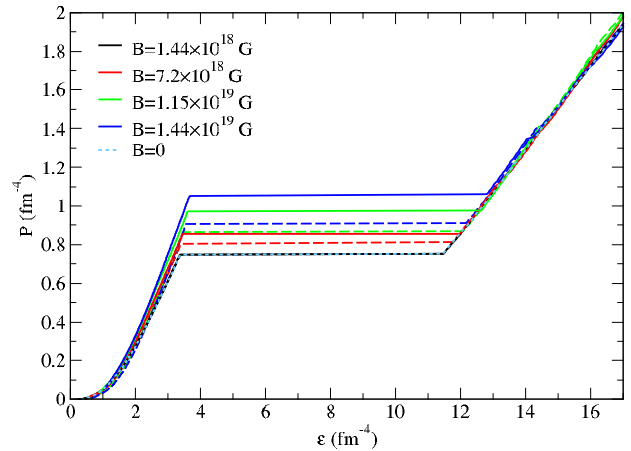}
\vspace{-1.cm}
\caption{{\bf{CMF model: }}EoS for neutron-star matter at $T=0$ for all constant magnetic-field strengths examined both with (solid) and without (dashed) AMM effects.}
\label{eosT0Star}
\end{figure}

We begin by discussing the equation of state (EoS), pressure $P$ vs. energy density $\epsilon$ for neutron-star matter at $T=0$ and all choices of constant magnetic field strength in Fig.~\ref{eosT0Star}. The EoS is shown both with (solid) and without (dashed) the effects of AMM. The most prominent feature of this figure is the presence of the first order phase transition between the hadronic and quark phases, as indicated by the horizontal lines (discontinuities in energy density) in the center of the figure. In the lower energy density region, we have a hadronic phase, while at higher energy densities we have a quark phase. For stronger magnetic fields, the phase transition takes place at slightly larger energy densities and the energy density gap between the end of the hadronic phase and the start of the quark phase increases significantly, i.e. the phase transition gets more pronounced {(stronger)}. The former was {already observed for the $T=0$ case for the CMF model in Ref.~\cite{Dexheimer_2012}, for a Walecka-type model combined with the MIT bag model in \cite{Rabhi:2009ih}, a Walecka-type model combined with the dependent quark mass 
model \cite{Backes:2021mdt}, a Walecka-type model combined with the Field Correlator Method model \cite{Mariani:2022xek}, a density-dependent model combined with the bag model \cite{Rather:2022bmm}, and in the Friedberg-Lee model in Ref.~\cite{Mao:2015kva}}.

\begin{figure}[t!]
\centering
\vspace{-.55cm}
\includegraphics[width=1.1\linewidth]{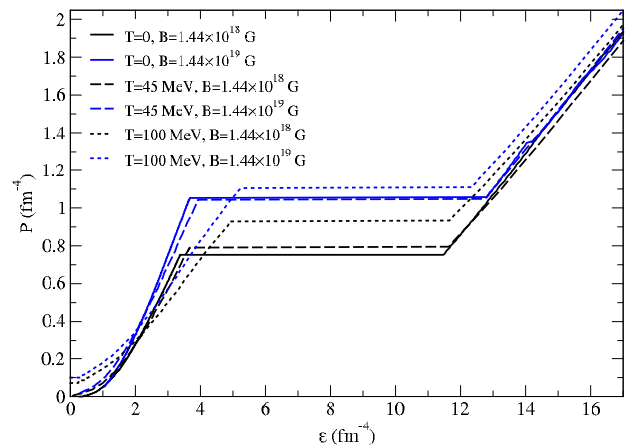}
\vspace{-1.cm}
\caption{{\bf{CMF model: }} {EoS for neutron-star matter at $T=0$ (full lines), 45 (dashed lines) and 100 (dotted lines) MeV} and for the strongest and weakest {(effectively zero)} nonzero magnetic fields with AMM effects examined in Fig.~1.}
\label{EOSallT_Star}
\end{figure}

\begin{figure}[t!]
\centering
\vspace{-.55cm}
\includegraphics[width=1.1\linewidth]{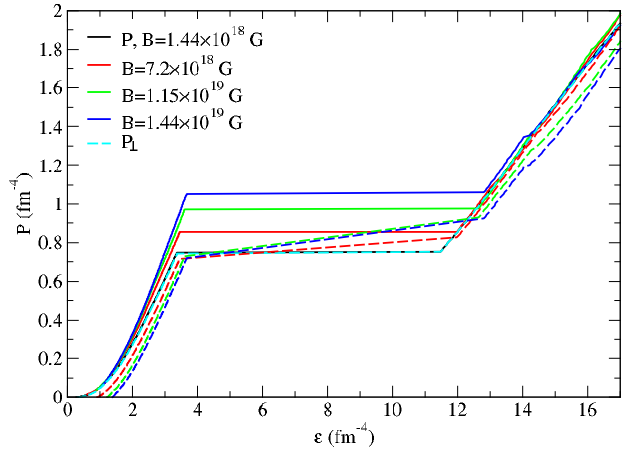}
\vspace{-1.cm}
\caption{{\bf{CMF model: }}Pressure (solid) and perpendicular pressure (dashed) as functions of energy density for neutron star matter at $T=0$ for all nonzero constant magnetic-field strengths examined with AMM effects. {The perpendicular pressure is discontinuous across the phase transition.}}
\label{8PT0Star}
\end{figure}

Additionally, we see in Fig.~\ref{eosT0Star} that increasing the magnetic field strength results in an overall stiffer EoS (larger $P$ for a given $\epsilon$), which would result in more massive neutron stars. {This result is model and density dependent, as shown in Fig.~2 of Ref.~\cite{Rabhi:2008je} and Fig.~8 of Ref.~\cite{Chu:2014pja}.} However, we also point to the presence of De Haas-Van Alphen (DHVA) oscillations \cite{DHVA}, whose behavior is related to the discrete nature of the Landau levels. They are more prominent for quark matter, due to their lower masses (when compared to baryons), and lead to EoS's that are softer (under stronger magnetic fields) than their weaker magnetic field counterparts for some energy densities. An additional (more conspicuous) softening of the EoS related to the appearance of strange quarks is visible in the quark phase; it appears as a cusp in the blue curve for $B=1.44\times10^{19}$ G. The inclusion of the AMM results in a stiffer EOS (as discussed in 
Ref.~\cite{broderick:2000pe}) with a stronger phase transition for the same magnetic field strength. The effect of including the AMM is comparable in magnitude to the magnetic field effects without the AMM: 
{just before the phase transition ($\varepsilon=3.38$ fm$^{-4}$) there is a $11\%$ pressure increase from $B=0$ to $B=1.44\times10^{19}$ G and an additional $11\%$ pressure increase when accounting for AMM in the latter case.} We also see that the cases of $B=0$ and $B=1.44\times10^{18}~G$ are indistinguishable from each other in the EoS. As it is known that, for a given value of magnetic field strength, the effects of magnetic fields diminish with increasing temperature \cite{Strickland:2012vu},  we do not need to consider the $B=0$ case at higher temperatures.

\begin{table}[t!]
\centering
\caption{{\bf{CMF model: }}Summary table showing the baryon chemical potential at the (quark deconfinement) phase transition and energy density at the beginning and end of the phase transition for neutron-star matter for all three temperatures and the strongest and weakest nonzero constant magnetic-field strengths analyzed {with AMM effects}. The subscripts $h$ and $q$ indicate whether the value is for the hadronic or quark side of the phase transition. The last column shows the energy density jump across the phase transition.}
\vspace{.5cm}
		\def\arraystretch{1.8}
		\begin{tabular}{|c|c|c|c|c|c|}
			\hline
		\ \ \ \ \ T\ \ \ \ \ & B  &\ \ \ \ \ $\mu_{B}$\ \ \ \ \   & \ \ \ \ \ $\epsilon_{h}$\ \ \ \ \  &\ \ \ \ \   $\epsilon_{q}$\ \ \ \ \   & \ \ \ \ $\Delta \epsilon$\ \ \ \   \vspace{-.2cm}\\
             (MeV)& (G)  &(MeV)  &(fm$^{-4}$) & (fm$^{-4}$) & (fm$^{-4}$)                \\
                                \hline
			0   & $1.44\times10^{18}$ & 1344 & 3.38 & 11.47 & 8.09 \\
			0   & $1.44\times10^{19}$ & 1368 & 3.68 & 12.81 & 9.14 \\
                                \hline
			45  & $1.44\times10^{18}$ & 1306 & 3.67 & 11.65 & 7.98 \\
			45  & $1.44\times10^{19}$ & 1323 & 3.93 & 12.75 & 8.82 \\
                                \hline
			100 & $1.44\times10^{18}$ & 1126 & 4.96 &  11.66 & 6.70 \\
			100 & $1.44\times10^{19}$ & 1135 & 5.26 & 12.33 & 7.07 \\
			\hline
		\end{tabular}
\label{StarDensity}
\end{table}

Fig.~\ref{EOSallT_Star} also shows EoS's, except now for the three temperatures investigated and only for $B=1.44\times10^{18}$ G (black) and $B=1.44\times10^{19}$ G (blue), both with AMM effects. This figure highlights the fact that the magnetic field strengths and temperatures we consider have similar effects in the EoS. At higher temperatures, the DHVA oscillations are no longer present, resulting in the EoS being stiffer for stronger magnetic fields at all energy densities. The phase transition at $T>0$ is still {very} prominent. As temperature increases, the phase transition takes place at larger energy densities and becomes less pronounced. The weakening of the phase transition results in a {smaller slope} of the EoS {on the hadronic side leading to a smaller jump in energy density} that is most prominent at $T=100$ MeV. To summarize the phase transition thresholds and strengths, we compare all cases discussed so far in Table \ref{StarDensity}, also indicating the baryon chemical potential $\mu_B$ at which the phase transition takes place. It clearly increases with magnetic-field strength and decreases with temperature. 

\begin{figure}[t!]
\centering
\vspace{-.55cm}
\includegraphics[width=1.1\linewidth]{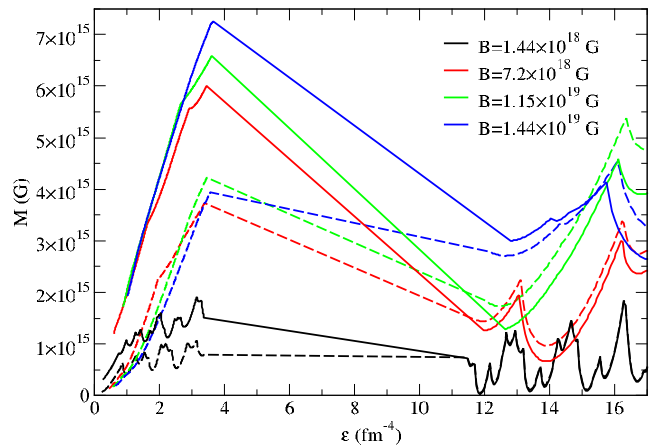}
\vspace{-1.cm}
\caption{{\bf{CMF model: }}Magnetization as a function of energy density for neutron-star matter at $T=0$ for all nonzero constant magnetic fields examined, both with (solid) and without (dashed) the effects of the AMM.}
\label{MagT0Star}
\end{figure}

\begin{figure}[t!]
\centering
\vspace{-.55cm}
\includegraphics[width=1.1\linewidth]{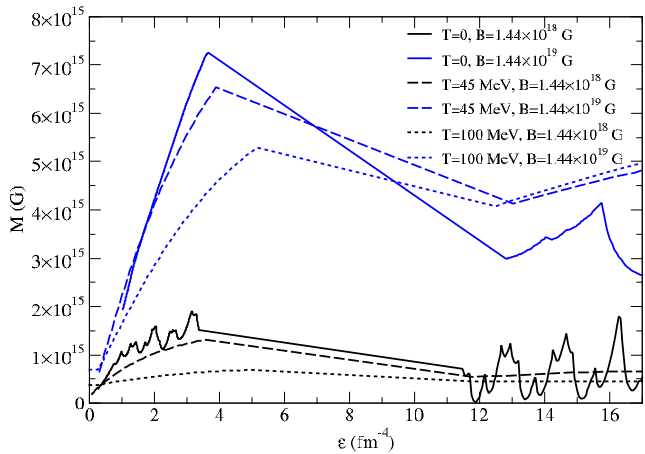}
\vspace{-1.cm}
\caption{{\bf{CMF model: }} {Magnetization as a function of energy density for neutron-star matter at $T=0$ (full lines), 45 (dashed lines) and 100 (dotted lines) MeV} and the strongest and weakest nonzero constant magnetic field strengths examined with AMM effects}.
\label{6MagStar}
\end{figure}

Fig.~\ref{8PT0Star} shows the pressure in the local direction of the magnetic field $P$, which is also referred to as parallel pressure (solid) and the pressure in the direction perpendicular to the field $P_\perp$ (dashed) as functions of energy density for $T=0$ and all magnetic-field strengths analyzed including AMM effects. In the latter case, the pressure receives a contribution from the magnetization
\begin{equation}
P_\perp=P-M B ,
\label{p-pperp}
\end{equation}
where the magnetization reflects how much the system is affected by the magnetic field $M=dP/dB$ (see Ref.~\cite{Strickland:2012vu} for a formal derivation of { Eq.~(\ref{p-pperp})} at zero and finite temperature including AMM effects). The perpendicular pressure differs from the pressure shown in the EoS in a few ways. First, stronger magnetic fields result in lower perpendicular pressures at the same energy density, the opposite of what is seen in the pressure. Second, the perpendicular pressure is negative for low energy densities $\lesssim1.5~fm^{-4}$. This is not physical, indicating that these ultra strong magnetic fields cannot exist at such low energy density. Realistic magnetic-field profiles for astrophysics will be discussed in the following and this is not an issue for heavy-ion collisions due to temperature contributions to the pressure. Finally, the perpendicular pressure is discontinuous over the phase transition. This discontinuity indicates that the magnetization will also be discontinuous over the phase transition. Not shown in this figure is that the size of the discontinuity in the perpendicular pressure decreases for higher temperatures and increases for stronger magnetic fields (the parallel pressure remains continuous in any case). However, this effect will be visible in the figures showing the magnetization.

 \begin{figure}[t!]
\centering
\vspace{-.55cm}
\includegraphics[width=1.09\linewidth]{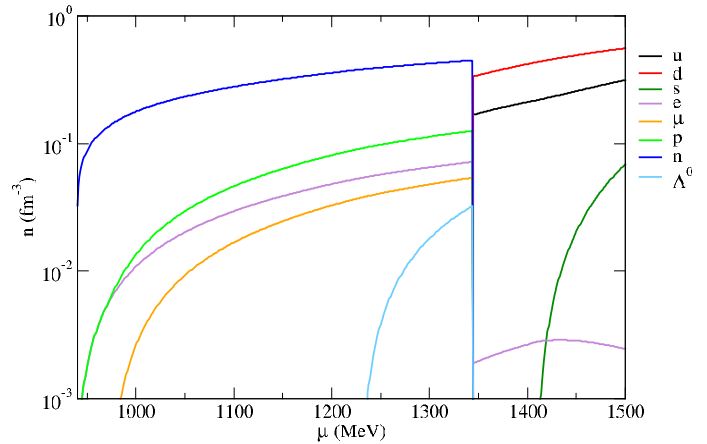}
\vspace{-1.cm}
\caption{{\bf{CMF model: }}Particle populations as functions of chemical potential for neutron-star matter at $T=0$ and $B=0$.}
\label{nmuT0B0Star}
\end{figure}

Fig.~\ref{MagT0Star} shows magnetization as a function of energy density at $T=0$ for several magnetic field strengths, both with (solid) and without (dashed) the effects of the AMM. Once again, the presence of the phase transition is clear {from the} jump in energy density. As expected, after examining the perpendicular pressure, the magnetization is discontinuous over the (first-order) phase transition. The presence of DHVA oscillations is much more clear in the magnetization than in the pressure or perpendicular pressure. As the magnetic field increases, the peaks and troughs of the magnetization oscillations tend to increase, as well as the width of the oscillations. In the hadronic phase, {the} inclusion of the AMM of baryons leads to larger magnetization, which is indicative of a larger pressure anisotropy. Double peaks indicate the different behavior of different spin projections. In the quark phase, the magnetization at $B=1.44\times10^{18}$ G is identical regardless of the AMM, while the stronger magnetic fields generally have a stronger magnetization without the AMM included. This difference comes from electrons, which are shown to have a nonzero population at higher magnetic fields {(shown} in Fig.~6 {and Fig.~7}).

Fig.~\ref{6MagStar} shows magnetization at several temperatures and $B=1.44\times10^{18}$ G and $B=1.44\times10^{19}$ G both with AMM effects. In the energy density range shown, as temperature increases, the magnetization decreases in overall magnitude and in the effect of DHVA oscillations, which is again expected, as the effect of the magnetic field becomes less pronounced at larger temperatures. These cannot be fully seen in the quark phase because of the range shown for the energy density. In addition, the magnetization discontinuity gap decreases with increased temperatures and increases with increased magnetic field strength.

 \begin{figure}[t!]
\centering
\vspace{-.55cm}
\includegraphics[width=1.09\linewidth]{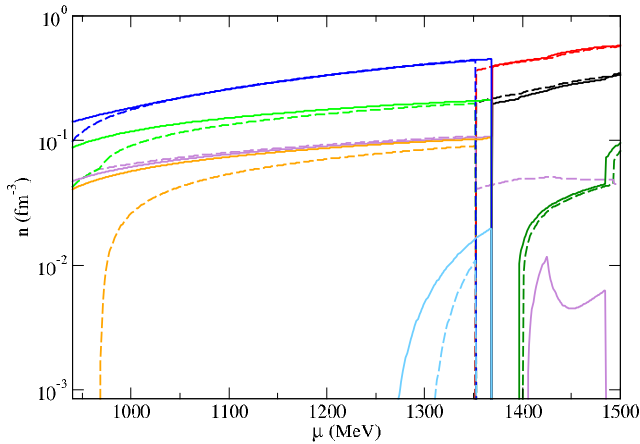}
\vspace{-1.cm}
\caption{{\bf{CMF model: }}Particle populations as functions of baryon chemical potential for neutron-star matter at $T=0$ and $B=1.44\times10^{19}$ G with (solid) and without (dashed) AMM effects. {See previous figure for labels.}}
\label{nmuT0B10E5Star}
\end{figure}

Fig.~\ref{nmuT0B0Star} shows particle populations as functions of baryon chemical potential (to avoid the gap in other {variables} across the phase transition) at $T=0$ and $B=0$. In this case, the phase transition occurs at $\mu_B=1344$ MeV. Leptons (electrons and muons) only appear {in significant amounts} in the hadronic phase, where neutrons are the most populous particle species. They are more than four times as populous as protons, the second most populous particle species. {Nevertheless}, both electrons and muons assist in achieving charge neutrality. Muons first appear at $\mu_B=970$ MeV and, by the phase transition, they account for approximately $40\%$ of the total proton charge, with electrons making up the difference. The only hyperon with nonzero population is $\Lambda^0$, which first appears for chemical potential $\mu_B=1229$ MeV, all others are suppressed by the phase transition. In the quark phase, down quarks are the most populous, being nearly twice as populous as up quarks until the more massive strange quarks begin to appear at $\mu_B=1408$ MeV. Notably, the total strangeness of the system changes at the phase transition.

Fig.~\ref{nmuT0B10E5Star} also shows particle populations as functions of baryon chemical potential, except now for the strongest magnetic field strength examined, $B=1.44\times10^{19}$ G, with (without) the inclusion of AMM  effects, shown in the full (dashed) lines. The phase transition occurs at $\mu_B=1368$ MeV ($\mu_B=1352$ MeV). Neutrons remain the most populous particles in the hadronic phase, though only by a factor of about $1.5-2.1$ over the protons, depending on the chemical potential. Electrons and muons remain the only means to achieve charge neutrality, but now they each account for close to $50\%$ of the total proton charge. Again, $\Lambda^0$ is the only hyperon to have a nonzero population. 

In the quark phase, there are {visible} DHVA oscillations in all three quark populations, with strange quarks seeing the largest impact from increases in Landau level. The down quark population is not strictly monotonic, occasionally decreasing slightly, indicating that some down quarks are changing flavor into strange quarks. The electrons in the quark phase are providing charge neutrality when there are too many up quarks for the combination of down and strange quarks to make up for as additional particles are blocked by Pauli exclusion. There are two drops in the quark phase electron population. The first coincides with a DHVA oscillation in the down quarks and the second with an oscillation in the strange quarks. Strange quarks first appear at $\mu_B=1397$ MeV, so there is again a drop to zero strangeness at the phase transition {(as in the $B=0$ case)}. 

\begin{figure}[t!]
\centering
\vspace{-.55cm}
\includegraphics[width=1.1\linewidth]{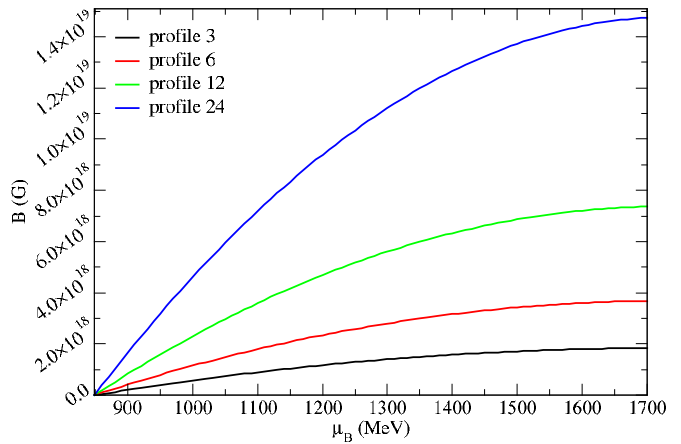}
\vspace{-1.cm}
\caption{Magnetic field profiles as functions of baryon chemical potential from Ref.~\cite{DEXHEIMER2017487}.}
\label{Bmu}
\end{figure}

Still discussing Fig.~\ref{nmuT0B10E5Star}, without the AMM, some particles do not appear until reaching a larger baryon chemical potential. In the hadronic phase, the muon and $\Lambda^0$ exhibit this behavior, while in the quark phase, it is seen in the strange quark. For {the muon and $\Lambda^0$}, this is due to the AMM term reducing the magnetic effective mass $\bar{m}_i$. The reduced mass allows the particles to exist in the system at baryon chemical potential lower than they otherwise could. At such a strong magnetic field, the electron mass is dominated by the AMM term except at very high Landau levels, leading to a suppression of electrons when the AMM is included, {visible in the quark phase. The differences in the quark phase exist only because of the electron AMM and imposed charge neutrality}.

\begin{figure}[t!]
\centering
\vspace{-.55cm}
\includegraphics[width=1.1\linewidth]{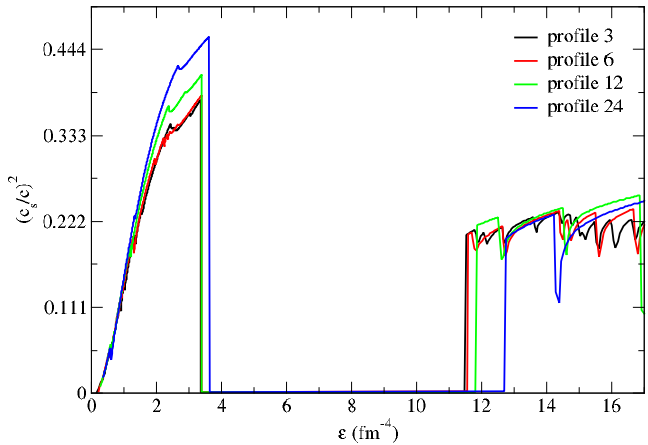}
\vspace{-1.cm}
\caption{{\bf{CMF model: }}Speed of sound squared as a function of energy density for neutron-star matter at $T=0$ for several examined magnetic field profiles with AMM effects.}
\label{soundT0prof}
\end{figure}

\begin{figure}[t!]
\centering
\vspace{-.55cm}
\includegraphics[width=1.1\linewidth]{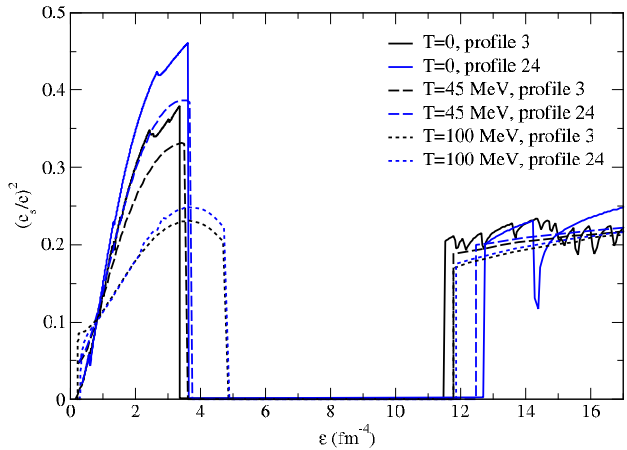}
\vspace{-1.cm}
\caption{{\bf{CMF model: }} {Speed of sound squared as a function of energy density for neutron-star matter at $T=0$ (full lines), 45 (dashed lines) and 100 (dotted lines) MeV } and magnetic field profiles 3 and 24 with AMM effects.}
\label{soundBprof}
\end{figure}

At finite temperature, the discussion of particle populations becomes more complicated, as all particles appear at all densities. For this reason, we do not show these figures. The most prominent effects are the more similar amounts of neutron and protons, of electrons and muons, and of $\Lambda^0$ and $\Sigma^-$ when the magnetic-field strength is large.

To model realistic neutron star interiors, we make use of the realistic magnetic field profile for neutron stars discussed in Section 
II A. Fig.~\ref{Bmu} shows the different magnetic-field profiles as a function of baryon chemical potential, going to a low value reached at the lowest stellar  core densities, until a large value beyond what is reached in the {center of neutron stars}.

In order not to repeat all our EoS results here with the magnetic-field profiles, we show instead the EoS derivatives, namely the speed of sound squared $c_s^2=dP/d\epsilon$. This quantity has been shown to be directly relevant for astrophysical discussions of, for instance, {stellar masses, radii, and tidal deformability \cite{Tan:2021ahl,Tan:2021nat,Ivanytskyi:2022bjc,Pinto:2022elo,Altiparmak:2022bke,Brandes:2022nxa}, and the outcome of binary neutron-star mergers \cite{Huang:2022mqp,Gorda:2022jvk,Jiang:2022tps}}. Besides, it has been suggested that this quantity may give some information about a possible deconfinement phase transition {\cite{Tan:2021nat}}. Within an agnostic description of the EOS and imposing constraints from perturbative QCD it was shown that it is expected that the speed of sound presents a pronounced peak around three times saturation density \cite{Annala:2019puf,Altiparmak:2022bke,Gorda:2022jvk,han2022plausible}, followed by a steep drop.
As expected, all {of our} curves show a drop to zero across the phase transition (see Fig.~\ref{soundT0prof}). This happens because, at the phase transition, the pressure is constant while the energy density jumps. Without magnetic field, spikes show the appearance of hyperons (in the hadronic phase) and strange quarks (in the quark phase). With magnetic fields,  we also see the presence of DHVA oscillations. Even the magnetic field profile 3, which in the EoS, shows no indication of magnetic effects (not shown {for the profiles}), shows clear DHVA oscillations in the speed of sound. As the magnetic field increases, the speed of sound squared generally increases and the DHVA oscillations grow in both amplitude and period, however, the strongest magnetic field, profile 24, has a lower speed of sound in the quark phase.

\begin{figure}[t!]
    \centering
        \vspace{-.55cm}
    \includegraphics[width=1.1\linewidth]{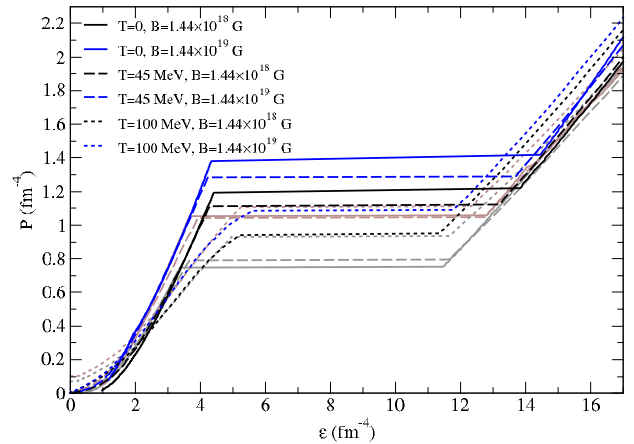}
    \vspace{-1.cm}
    \caption{{\bf{CMF model: }} {EoS for heavy-ion collision matter for $T=0$ (full lines), 45 (dashed lines) and 100 (dotted lines) MeV } and the strongest and weakest constant magnetic fields strengths examined with AMM. The gray corresponds to the black lines and the brown to the blue lines in the case of neutron star matter, for comparison.}
    \label{EOSallT_Sym}
\end{figure}

Fig.~\ref{soundBprof} is the same as Fig.~\ref{soundT0prof}, except now shown for several temperatures and only magnetic field profiles 3 and 24 with AMM effects. The DHVA oscillations are suppressed by the temperature effects and, as a result, a stronger magnetic field results in a larger speed of sound. Also, the speed of sound begins to decrease in the hadronic phase prior to the phase transition, this is more evident for $T=100$ MeV than for $T=45$ MeV and it is a result of quarks starting to appear inside the hadronic phase as the phase transition becomes weaker, which means that we are approaching the critical point for deconfinement. In the quark phase for all temperatures, we note that the speed of sound squared stays below the conformal limit $(c_s/c)^2\le1/3$, which is expected from perturbative QCD calculations \cite{Fraga:2013qra}.

Now, we change our discussion to matter produced in heavy-ion collisions, as described in the beginning of this Section.
Fig.~\ref{EOSallT_Sym} shows EoS for all temperatures and the strongest and weakest magnetic field strengths with AMM. The gray and brown curves are a repetition of the neutron star EoS in Fig.~\ref{EOSallT_Star}. Gray/brown corresponds to the same temperature and magnetic field as black/blue for heavy-ion collisions matter.
Heavy-ion matter always reaches a higher energy density prior to the phase transition and is {overall stiffer than neutron-star matter}, except just prior to the phase transition at $T=100$ MeV (discussed below). The stiffer EoS for heavy-ion matter is due to Pauli exclusion, as there are no {hyperons in zero net strangeness isospin-symmetric matter (a larger effect than the one related to the asymmetry between protons and neutrons)}. For heavy-ion matter, larger magnetic field strength only corresponds to a phase transition at larger energy densities (as in the neutron-star case) for the $T=100$ MeV case. For the lower temperatures, the behavior is the opposite.
The {softer curve (with a smaller slope)} of the hadronic phase toward the phase transition at $T=100$ MeV (already discussed for neutron-star matter) is also present for isospin symmetric matter, but now  it is also more discernible, {indicating} a weaker phase transition. 

\begin{figure}[t!]
    \centering
        \vspace{-.55cm}
    \includegraphics[width=1.1\linewidth]{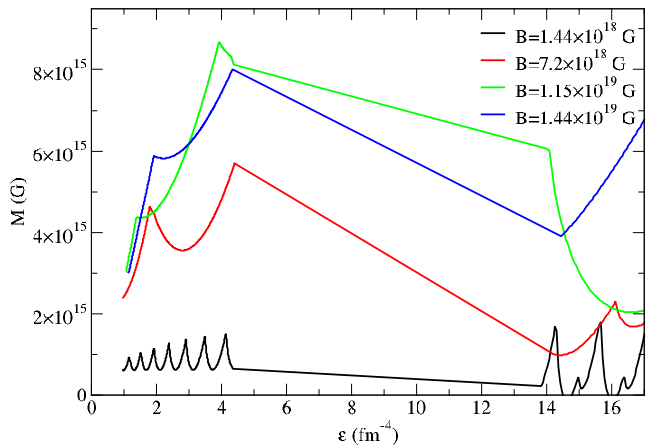}
    \vspace{-1.cm}
    \caption{{\bf{CMF model: }}Magnetization as a function of energy density for symmetric matter at $T=0$ for all nonzero magnetic fields examined with AMM.}
    \label{MagT0Sym}
\end{figure}

Furthermore,  for $T=0$ and $T=45$ MeV, heavy-ion matter presents a slightly larger jump in energy density (stronger phase transition), whereas at $T=100$ MeV, the phase transition is weaker when compared to neutron-star matter. This means that the critical point (when there is no longer a discontinuity in the energy density, and thus, the phase transition ceases to be first order) of the deconfinement phase transition will occur at a lower temperature for heavy-ion matter than for neutron-star matter. This has been previously discussed within the CMF model at $B=0$ in Ref. \cite{Aryal2021}. The exact values of the differences in energy density between the beginning and end of the phase transition are given in Table \ref{energySymStar}. 

 \begin{figure}[t!]
    \centering
        \vspace{-.55cm}
    \includegraphics[width=1.1\linewidth]{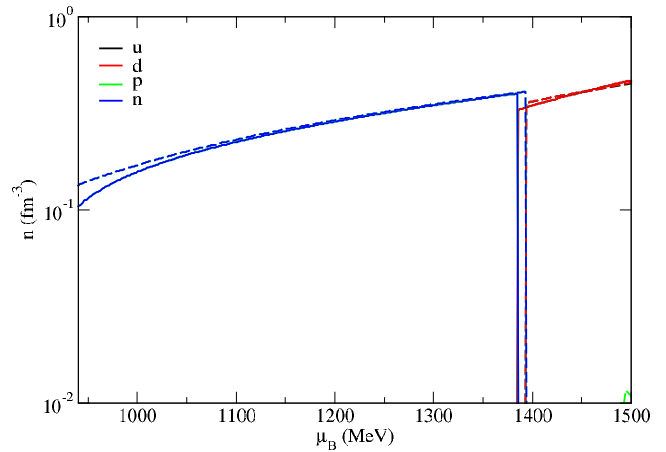}
    \vspace{-1.cm}
    \caption{{\bf{CMF model: }}Particle populations as functions of chemical potential for heavy-ion collisions matter at $T=0$ and the weakest (solid) and strongest (dashed) nonzero constant magnetic fields examined with AMM {many curves overlap}.}
    \label{nmuT0Sym}
\end{figure}

\begin{table}[t!]
\centering
\caption{{\bf{CMF model: }}Summary table showing the change in energy density across the phase transition for neutron-star matter and heavy-ion collisions matter and which type of matter has the stronger deconfinement phase transition for several temperatures and the strongest and weakest constant magnetic field strengths with AMM.}
		\def\arraystretch{1.8}
		\begin{tabular}{|c|c|c|c|c|}
			\hline
			\ \ \ \ \ \ T\ \ \ \ \ \ & B  & \ \ \ $\Delta\epsilon_{NS}$\ \ \   & $\ \ \Delta\epsilon_{HIC}$\ \   & \ \ \ \ Str.\ \ \ \ \vspace{-.2cm}\\
             (MeV)& (G)&  (fm$^{-4}$)    &(fm$^{-4}$)& PT                       \\
                                \hline
			0     & $1.44\times10^{18}$ & 8.09 & 9.42   & HIC \\
			0     & $1.44\times10^{19}$ & 9.14 & 10.11 & HIC \\
                                \hline
			45   & $1.44\times10^{18}$ & 8.18 & 9.09   & HIC \\
			45   & $1.44\times10^{19}$ & 9.14 & 9.78   & HIC \\
                                \hline
			100 & $1.44\times10^{18}$ & 6.94 & 6.44   & NS \\
			100 & $1.44\times10^{19}$ & 7.35   & 6.58   & NS \\
			\hline
		\end{tabular}
\label{energySymStar}
\end{table}

Fig.~\ref{MagT0Sym} shows magnetization as a function of energy density at $T=0$ for heavy-ion matter. There are many similarities with Fig.~\ref{MagT0Star} but, unlike for neutron-star matter, the magnetization goes negative for some energy densities in the quark phase for $B=1.44\times10^{18}$ G, indicating a slight diamagnetic behavior.

To finalize, we discuss once more the particle populations, but now for the case of heavy-ion matter. Fig.~\ref{nmuT0Sym} shows particle populations as functions of baryon chemical potential for $T=0$ and $B=1.44\times10^{18}$ G (solid) and $B=1.44\times10^{19}$ G (dashed). The phase transitions occur at $\mu_B=1385$ MeV for the weaker magnetic field and $\mu_B=1393$ MeV for the stronger magnetic field. As expected due to the constraint of zero net strangeness, in the hadronic phase, neutrons and protons are the only particles present and {appear to} have equal populations. In the quark phase, up and down quarks are the only particles present and also {appear to} have equal populations. For larger magnetic field strength, there are more particles (for a given chemical potential). 

For the stronger magnetic field, there is also a very small splitting between proton and neutron and between the up and down populations, which becomes larger for larger temperatures (not shown here). This is due to fact that isospin symmetry {was defined in this work  
 by setting the isospin chemical potential $\mu_I$ to zero, in order to make the chemical potential of particles that differ by isospin equal, $\mu_p=\mu_n$ and $\mu_u=\mu_d$. At zero magnetic field, this results in equal populations of protons and neutrons and up and down quarks.} However, for a strong magnetic field, this is not the case because of the way the magnetic field influences the effective mass of charged {and} uncharged particles (due to {both the AMM and another charge dependent term \cite{Strickland:2012vu}}), and consequently their momenta and density. {A different way to address this issue would be to impose directly equal densities $n_p=n_n$  and $n_u=n_d$, which would necessarily imply $\mu_p\neq\mu_n$ and $\mu_u\neq\mu_d$.}

\subsection{PNJL Model}

\begin{figure}[t!]
    \centering
        \vspace{-.55cm}
    \includegraphics[width=1.1\linewidth]{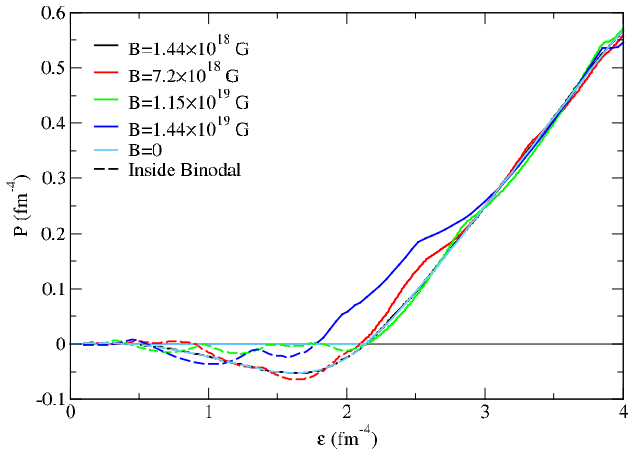}
    \vspace{-1.cm}
    \caption{{\bf{PNJL model: }}EoS for neutron-star matter at $T=0$ for all constant magnetic-field strengths.}
    \label{PNJL1}
\end{figure}

For comparison with the CMF model, we present similar figures now for the PNJL model. We start with the $T = 0$ neutron-star matter EoS for all constant magnetic-field strengths studied in the previous Subsection. The AMM is not included in the PNJL model, as it is not clear if it is relevant for quarks \cite{Weinberg:1990xm,Ferrer:2015wca} (it was not included for quarks in the CMF model). Once more, the most prominent feature of Fig.~\ref{PNJL1} is the presence of the first-order phase transition, indicated by horizontal lines between the constituent quark phase (on the left), where chiral symmetry is broken, and quark {phase} (on the right), where chiral symmetry is already partially restored. These can be interpreted as hadronic and quark phases. {Additionally, we now also show the meta-stable and unstable phases as dashed lines. The metastable regions, which can present negative pressure, define the binodal, where a Maxwell construction is applied. The unstable regions present a negative slope in pressure and define the spinodal. Although metastable regions are not necessarily relevant for mechanically-equilibrated neutron stars, they are very important for heavy-ion collisions, due to the very short time scales involved.}

At $T=0$, independently of the magnetic field strength, the phase transition starts at $\epsilon\sim 0$. The end depends on the magnetic field, but it is not clear if it pushes the phase transition to larger or lower energy densities and if the size of the gap increases or decreases. This is related to the fact that the DHVA oscillations are very strong for quark matter, as already discussed {for the CMF model}. 

\begin{figure}[t!]
    \centering
        \vspace{-.55cm}
    \includegraphics[width=1.1\linewidth]{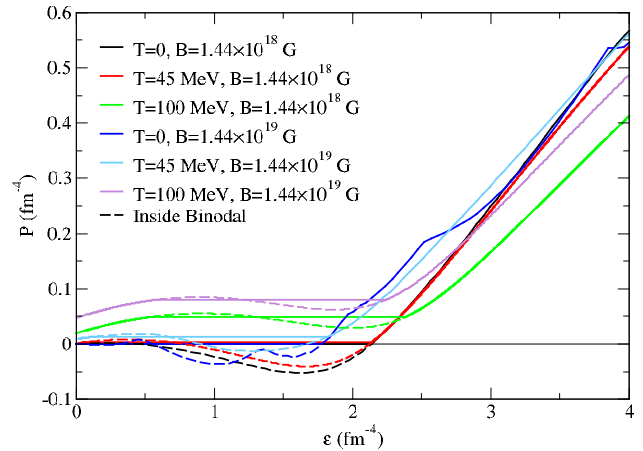}
    \vspace{-1.cm}
    \caption{{\bf{PNJL model: }}EoS for neutron-star matter {for $T=0$ (full lines), 45 (dashed lines) and 100 (dotted lines) MeV } and for the strongest and weakest nonzero magnetic fields examined in Fig.~14.}
    \label{PNJL2}
\end{figure}

\begin{table}[t!]
\centering
\caption{{\bf{PNJL model: }}Summary table showing the baryon chemical potential at the (quark deconfinement) phase transition and energy density at the beginning and end of the phase transition for neutron-star matter for all three temperatures and the strongest and weakest nonzero constant magnetic-field strengths analyzed with AMM. The subscripts $h$ and $q$ indicate whether the value is for the hadronic or quark side of the phase transition. The last column shows the energy density jump across the phase transition.}
\vspace{.5cm}
		\def\arraystretch{1.8}
		\begin{tabular}{|c|c|c|c|c|c|}
			\hline
				\ \ \ \ \ T\ \ \ \ \ & B  &\ \ \ \ \ $\mu_{B}$\ \ \ \ \   & \ \ \ \ \ $\epsilon_{h}$\ \ \ \ \  &\ \ \ \ \   $\epsilon_{q}$\ \ \ \ \   & \ \ \ \ $\Delta \epsilon$\ \ \ \   \vspace{-.2cm}\\
             (MeV)& (G)  &(MeV)  &(fm$^{-4}$) & (fm$^{-4}$) & (fm$^{-4}$)                \\
                                \hline
			0   & $1.44\times10^{18}$ & 1102 & $\approx$ 0 & 2.14 & 2.14 \\
			0   & $1.44\times10^{19}$ & 1075 & $\approx$ 0 & 1.76 & 1.76 \\
                                \hline
			45  & $1.44\times10^{18}$ & 1092 & 0.09 & 2.14 & 2.05 \\
			45  & $1.44\times10^{19}$ & 1068 & 0.14 & 1.82 & 1.68 \\
                                \hline
			100 & $1.44\times10^{18}$ & 1042 & 0.57 & 2.38 & 1.81 \\
			100 & $1.44\times10^{19}$ & 1026 & 0.59 & 2.24 & 1.65 \\
			\hline
		\end{tabular}
\label{StarDensityPNJL}
\end{table}

\begin{figure}[t!]
    \centering
        \vspace{-.55cm}
    \includegraphics[width=1.1\linewidth]{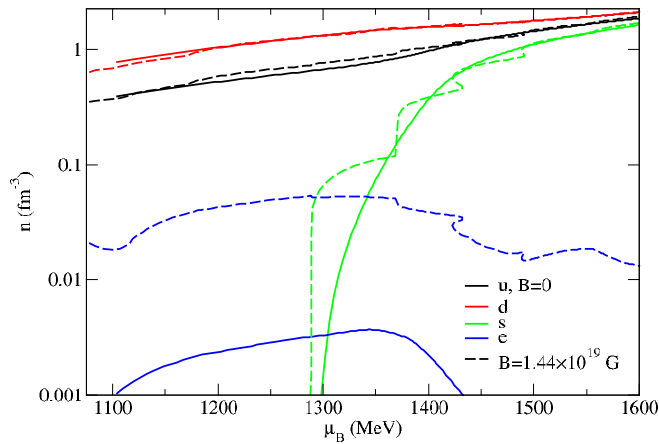}
    \vspace{-1.cm}
    \caption{{\bf{PNJL model: }}Particle populations {in the quark phase} as functions of baryon chemical potential for neutron-star matter at $T=0$ for $B=0$ (solid) and the highest constant magnetic field strength examined (dashed).}
    \label{PNJL67}
\end{figure}

Fig.~\ref{PNJL2} compares only the lowest and highest magnetic fields studied. In this case, for all temperatures studied, stronger magnetic fields push the quark side of the phase transition to lower energy densities and the size of the gap decreases. {From Table \ref{StarDensityPNJL}, at $T=0$ and finite temperature, we also see a consistent decrease of the critical baryon chemical potential as the magnetic field increases (from $B=0$ to the strongest value analyzed), resulting in the expected inverse  magnetic catalysis  at finite $\mu_B$ \cite{Avancini:2012ee,Ruggieri:2014bqa,Costa:2015bza,Ferreira:2013tba,Ferreira:2014kpa} (where the magnetic field enhances chiral symmetry). This is the opposite behavior of the CMF model.}

{Still discussing Fig.~\ref{PNJL2},} at larger temperatures, the DHVA oscillations are no longer present and the effects of magnetic fields diminish in the quark phase, resulting in the EoS being stiffer for stronger magnetic fields at all energy densities. The phase transition at $T>0$ is still prominent and, as temperature increases, the phase transition happens at larger energy densities, {lower baryon chemical potentials} and gets weaker, as expected, {and as already discussed for the CMF model}.

\begin{figure}[t!]
    \centering
        \vspace{-.55cm}
    \includegraphics[width=1.1\linewidth]{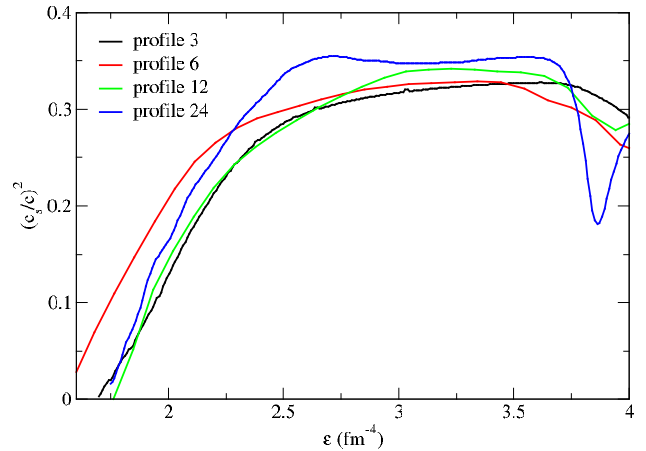}
    \vspace{-1.cm}
    \caption{{\bf{PNJL model: }}Speed of sound squared {in the quark phase} as a function of energy density for neutron-star matter at $T=0$ for several examined magnetic field profiles.}
    \label{PNJL9}
\end{figure}

\begin{figure}[t!]
    \centering
        \vspace{-.55cm}
    \includegraphics[width=1.1\linewidth]{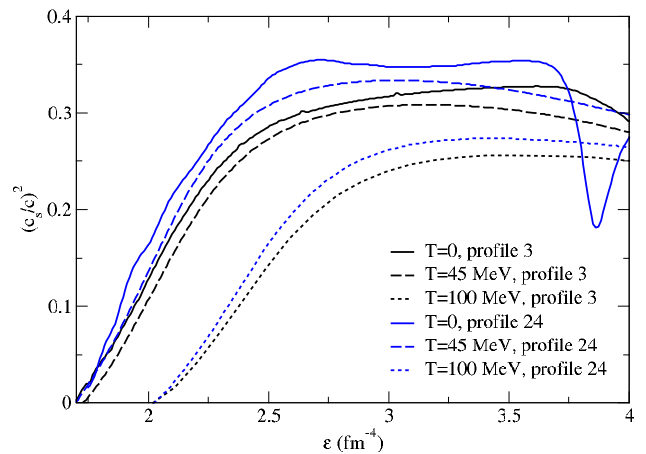}
    \vspace{-1.cm}
    \caption{{\bf{PNJL model:}} {Speed of sound squared {in the quark phase} as a function of energy density for neutron-star matter for $T=0$ (full lines), 45 (dashed lines) and 100 (dotted lines) MeV } and magnetic field profiles 3 and 24.}
    \label{PNJL10}
\end{figure}

Fig.~\ref{PNJL67} shows particle populations as functions of baryon chemical potential at $T=0$ for $B=0$ and $B=1.44\times 10^{19}$ G starting beyond where the phase transition takes place $\mu_B\sim1100$ MeV ({$\mu_B\sim1102$} for $B=0$ and $\mu_B\sim1075$ for $B=1.44\times 10^{19}$ G, see Table \ref{StarDensityPNJL}), showing only the quark phase. The only leptons included are the electrons. At $B=0$, they appear in very small amounts. The down quarks are about two times more populous than the up quarks, until the strange quarks appear. This happens at $\mu_B\sim1300$. For the strong magnetic field (shown in dashed lines), the phase transition takes place a bit earlier. In this case, there are more than ten times more electrons and overall more up and strange quarks. The DHVA oscillations can be seen in all populations.

Once more, to model realistic neutron star interiors, we make use of the magnetic field profile for neutron stars discussed in Section II A and shown in Fig.~\ref{Bmu}. In order not to repeat all our {PNJL} EoS results here with the magnetic-field profiles, we show instead {the} speed of sound squared, in the quark phase. In Fig.~\ref{PNJL9} at T=0, we see the presence of DHVA oscillations for all profiles (as for the CMF model). {Larger profiles (corresponding to larger magnetic field strengths) present higher bumps in speed of sound.} {The overall bump structure is related to the appearance of {strange} quarks, see Ref.~\cite{Tan:2021ahl} for a review on the discussion of structure in the speed of sound.}

Fig.~\ref{PNJL10} is the same as Fig.\ref{PNJL9}, except now shown for several temperatures and only magnetic field profiles 3 and 24. The DHVA oscillations are suppressed by the temperature effects and stronger magnetic fields (for $T>0$) result in larger speeds of sound.  For all temperatures, we note that the speed of sound squared for large energy densities stays below the conformal limit $(c_s/c)^2\le1/3$, {although there are {additional bump regions that oscillate} above the limit for $T=0$ (not shown here).}

\begin{figure}[t!]
    \centering
        \vspace{-.55cm}
    \includegraphics[width=1.1\linewidth]{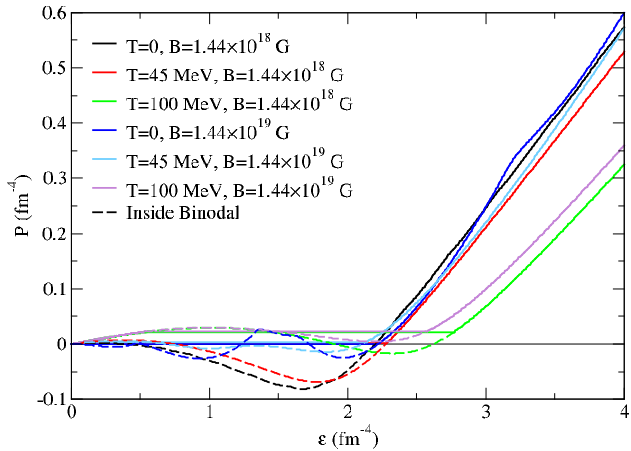}
    \vspace{-1.cm}
    \caption{{\bf{PNJL model: }}EoS for heavy-ion collision matter for {$T=0$, 45 and 100~MeV } and the strongest and weakest constant magnetic fields strengths examined.}
    \label{PNJL11}
\end{figure}

\begin{table}[t!]
\centering
\caption{{\bf{PNJL model: }}Summary table showing the change in energy density across the phase transition for neutron-star matter and heavy-ion collisions matter and which type of matter has the stronger deconfinement phase transition for several temperatures and the strongest and weakest constant magnetic field strengths.}
		\def\arraystretch{1.8}
		\begin{tabular}{|c|c|c|c|c|}
			\hline
					\ \ \ \ \ \ T\ \ \ \ \ \ & B  & \ \ \ $\Delta\epsilon_{NS}$\ \ \   & $\ \ \Delta\epsilon_{HIC}$\ \   & \ \ \ \ Str.\ \ \ \ \vspace{-.2cm}\\
             (MeV)& (G)&  (fm$^{-4}$)    &(fm$^{-4}$)& PT                       \\
                                \hline
			0     & $1.44\times10^{18}$ & 2.14 & 2.20   & HIC \\
			0     & $1.44\times10^{19}$ & 1.78 & 2.22 & HIC \\
                                \hline
			45   & $1.44\times10^{18}$ & 2.05 & 2.22   & HIC \\
			45   & $1.44\times10^{19}$ & 1.68 & 2.03   & HIC \\
                                \hline
			100 & $1.44\times10^{18}$ & 1.81 & 2.23   & HIC \\
			100 & $1.44\times10^{19}$ & 1.65   & 2.02   & HIC \\
			\hline
		\end{tabular}
\label{energySymStarPNJL}
\end{table}

Now, we once more change our discussion to matter produced in heavy-ion collisions. Fig.~\ref{PNJL11} shows the EoS for all temperatures and the strongest and weakest magnetic field strengths. {At $T=0$ the transition starts again at $\epsilon\sim0$ and ends slightly at lower $\epsilon$ in the presence of strong magnetic fields. At larger temperatures, the phase transition starts later, and ends later (in $\epsilon$), but still happens at lower $\epsilon$ for strong magnetic fields (unlike in the CMF model). The effect of magnetic fields on the jump in $\epsilon$ across the phase transition is not clear, but it decreases with chemical potential and temperature}.

For a fixed magnetic field, temperature effects are opposite between the quark and constituent quark phases. In the constituent quark phase (when it exists), increased temperature results in a stiffer EoS, whereas the EoS is softer at higher temperatures in the quark phase. Stronger magnetic fields at the same temperature result in a stiffer EoS, except when there are DHVA oscillations.
See Tab.~\ref{energySymStarPNJL} for details. 

\begin{figure}[t!]
    \centering
        \vspace{-.55cm}
    \includegraphics[width=1.1\linewidth]{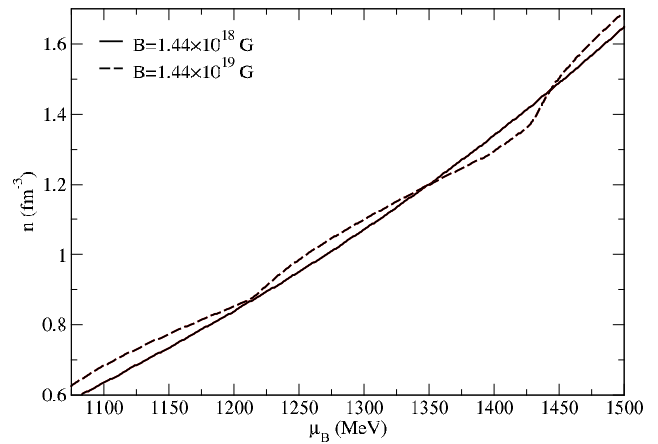}
    \vspace{-1.cm}
    \caption{{\bf{PNJL model: }}Particle populations {in the quark phase} as functions of chemical potential for heavy-ion collision matter at $T=0$ and the weakest (solid) and strongest (dashed) nonzero constant magnetic fields examined.}
    \label{PNJL13}
\end{figure}

Finally, we discuss the particle populations for the case of heavy-ion matter in the quark phase. Fig.~\ref{PNJL13} shows particle populations as functions of baryon chemical potential for $T=0$ {with} $B=1.44\times10^{18}$ G (solid) and $B=1.44\times10^{19}$ G (dashed). Due to the constraint of zero net strangeness, up and down quarks are the only particles present and also have equal populations. {In this case (unlike for the CMF model), there are no AMM effects and the isospin symmetry is fixed by fixing $n_u=n_d$ directly. For these reasons, the two curves overlap exactly.} For the larger magnetic field strength, there are generally more particles (for a given chemical potential) and {the} DHVA oscillations {are pronounced.} 

\section{Discussion and Conclusions}

With this work we aimed to provide an in depth analysis of how magnetic {fields} affect matter at extreme conditions. Such studies have become even more relevant in light of recent observations of neutron star merger events.  Giving continuity to foundation laid by previous works, we self-consistently included finite temperature effects and calculated the properties of matter at {large densities and under the influence of strong} magnetic fields. By appropriately choosing conditions {related to electric charge, isospin and strangeness conservation, as well as including leptons and considering different magnetic field configurations,} we are able to calculate microscopic properties and the equation of state for {astrophysical} conditions {and conditions produced in the laboratory}. 

{We focus on deconfinement to quark matter, that is expected to take place at large density. We then analyze how both temperature and strong magnetic fields can affect deconfinement and how conditions found in neutron stars and heavy-ion collisions change that. With this, we can estimate how both temperature and strong magnetic fields can affect deconfinement in neutron star mergers, which are expected to produce conditions that approach both neutron stars (in terms of charge/isospin \cite{Most:2018eaw} and magnetic fields \cite{Ciolfi:2013dta,Giacomazzo:2014qba,Ciolfi:2019fie,Most:2019kfe,Ciolfi:2020cpf,Palenzuela:2021gdo,Ruiz:2021qmm,Suvorov:2021ymy}) and heavy-ion collisions (with respect to temperature and entropy \cite{Most:2022wgo} {and magnetic fields \cite{Deng:2012pc,Taghavi:2013ena,Tuchin:2013apa}}), while producing an unprecedented amount of net strangeness and densities \cite{Most:2018eaw}.}

To study dense matter, we made use of two different models, namely the {Chiral Mean Field (CMF) and the Polyakov-loop extended Nambu–Jona-Lasinio (PNJL) - both of which are relativistic SU(3) chiral models}. Furthermore, these models present the advantage of realistically {and self-consistently describing both chiral symmetry restoration and deconfinement to quark matter. One main difference is that, while the CMF model includes baryons and quarks (and leptons), the PNJL model only includes quarks (and leptons), although it also presents a phase that mimics the hadronic one. Both models were fitted to reproduce low-energy nuclear physics, astrophysics, and lattice QCD in the regimes where they apply.

In order to describe the conditions of heavy-ion collisions, we calculate matter properties with {constant strong, but still realistic, magnetic field strengths. In this case,} anisotropy effects are {not expected to be relevant} in the {small size and} short time span in which extreme matter is created {during and} after the collision. For neutron star matter, however, we take a somewhat more sophisticated approach. Although the inner morphology of magnetic fields inside neutron stars is currently unknown, one expects that it should not be constant inside the neutron star, as a few works with consistent general relativistic calculations have shown {\cite{Bocquet:1995je,Cardall:2000bs,Frieben:2012dz,Pili:2014npa,Tsokaros:2021pkh}}. For that reason we modeled neutron star matter following a profile \cite{Dexheimer:2016yqu} fitted after general relativistic calculations, from which we were able to obtain a somewhat accurate description of the magnetic field as a function of the chemical potential. We note however that in this work we only use the polar {stellar} direction - a more complete study, fully considering all possible directions, thus providing a 2D map is currently ongoing.

The temperatures and magnetic fields, {as well as the anomalous magnetic moment, AMM (included only in the CMF model for hadrons and leptons)}, we study in this work have comparable effects on matter. Magnetic fields turn the EoS stiffer in both models, although at $T=0$ the DHVA oscillations (related to the quantization of energy into Landau levels) generate wiggles. These {wiggles} are much better seen in derivatives of the EoS, such as the speed of sound, where a zigzag pattern emerges. Note that some of the structure that appears in the speed of sound is related to new strange degrees of freedom (hyperons and strange quarks) appearing. For both models the speed of sound stays within the conformal limit at large densities (for the values analyzed).

In both models the temperature, as expected, pulls the phase transition to lower chemical potentials and weakens it (smaller jump in energy density across). This is quite natural, as both models predict critical points, beyond which the first-order deconfinement phase transition becomes a smooth crossover. This is not modified by the magnetic field. 
Discussion about the effect of strong magnetic fields on the critical point of the CMF model will be addressed in future work. For the PNJL, the phase transition starts at $\epsilon\sim0$ at $T=0$.
The PNJL model may also be coupled to a hadronic model at low densities to make the description of the low density EOS more realistic.
The successful description of two solar mass stars can then be a filter that indicates how large the jump in energy density can be (a jump that is too large may turn hybrid stars unstable \cite{Alford:2015gna}). 
The magnetic field makes the phase transition stronger for the CMF model and weaker for the PNJL model, at least in the studied range, presenting, respectively, a larger and smaller jump in energy density.This indicates that there is no universal behavior with respect to these quantities, and the outcome depends on the characteristics of the model.

For the CMF model, we also show for neutron-star matter the pressure locally perpendicular to the magnetic field, modified by the magnetization. It becomes softer for larger magnetic fields and presents a discontinuity across the phase transition. This discontinuity was addressed in Ref.~\cite{Ferrer:2020tlz} at zero temperature. Nevertheless, the true nature of the coexistence between the hadronic and quark phases under gravitational forces (and the possible appearance of mixtures of phases) depends strongly on the debated value of the surface tension \cite{Lugones:2018qgu} and a better understanding on how Landau levels behave at interfaces \cite{Chen:2017xrj}, a problem that depends on the geometry of the problem, which requires general relativity input. This goes beyond the scope of our work. The magnetization increases in value with magnetic field strength and becomes more smooth for larger temperatures. Double peaks point to different behavior for different spin projections.

Concerning particle populations, the magnetic field enhances charged particles, {(the larger the charge, the larger the enhancement)} turning the system {for neutron star matter} more isospin symmetric, and suppressing hyperons in the CMF model (the neutral $\Lambda$-hyperons {that appear in}  neutron-star matter). The leptons are enhanced for neutron-star matter, especially in the quark phase, where in the CMF model they can have additional effects due to the AMM effect. At large temperatures, the comparison becomes more complicated, as all particles appear at all densities.

Our calculations demonstrate that the different conditions in neutron star and heavy-ion collision matter -- namely the isospin symmetry and zero net strangeness found in heavy-ion collision matter, as opposed to charge neutrality and {weak} chemical equilibrium needed for neutron star matter -- can alter the effects of {magnetic field and temperature in dense} matter.

For larger magnetic fields, the deconfinement phase transition is pushed to larger energy densities and larger chemical potentials within the CMF model for neutron-star matter. This effect persists at all temperatures. For heavy-ion matter, this is only the case for $T=100$ MeV, and is opposite for lower temperatures (with respect to energy density). The phase transition, is stronger for the heavy-ion case for $T=0$ and $T=45$ MeV, but weaker for $T=100$ MeV.

In the PNJL model, the effect of strong magnetic fields is also not clear with respect to the energy density, but for the studied magnetic fields, their effect is to pull the chiral phase transition in an opposite manner than the CMF
model, to lower chemical potentials, an effect already identified as inverse catalysis at finite $\mu_B$.
Note that, although comparing these two models is the best we can do at the moment (there are no other models that provide a self-consistent description of deconfinement accounting for magnetic field effects), these models are quite different. The CMF is strongly affected by baryons and their AMM corrections; neither of which are included in the PNJL model.

For the CMF model, heavy-ion matter is overall stiffer, reaches a higher energy density prior to the phase transition, and presents a stronger phase transition (larger jump in energy density) than neutron-star matter {(containing hyperons)} for the same conditions of magnetic field and temperature. An exception is the case of $T = 100$ MeV, where heavy-ion matter matter becomes softer just prior to the phase transition and the phase transition is weaker. This is related to the proximity of the critical point, shown to appear at much lower temperatures for heavy-ion matter in the CMF model \cite{Aryal2021} (without magnetic-field effects).
For the PNJL model, heavy-ion matter is overall softer, reaches a lower energy density prior to the phase transition and a higher energy density after, and presents a stronger phase transition (compared to neutron star matter) in all temperature and magnetic field cases analyzed.

{An interesting related topic that has not been addressed in this work is the chiral magnetic effect \cite{Kharzeev:2007jp}, a generation of electric current induced by chirality imbalance in the presence of a magnetic field. There is indication that it has already been observed in both RHIC \cite{STAR:2015wza} and LHC \cite{Belmont:2014lta}. We note that, although we study the effect of strong magnetic fields in chiral models, we cannot addressed such effect in our work due to our assumption of space and time homogeneity, consequence of the mean-field approximation taken in both CMF and PNJL models.}

To summarize, we believe we have made great strides towards understanding the properties of dense matter at extreme conditions by considering {strong} magnetic fields and finite temperature on the same framework, obtaining thus a better understanding of matter in relevant scenarios such as neutron star mergers and heavy-ion collisions. However, there is still much to do and to learn. We aim, in a future work, to employ the knowledge gained by this study - to self-consistently model neutron stars with finite temperature and {strong} magnetic fields - both in the micro and macroscopic realms, using anisotropic solutions of Einstein and Maxwell equations.

In the next few years, we expect a very large amount of data  constraining dense matter, not only at $T\sim0$, but also at significantly larger temperatures, once gravitational wave interferometers measure the post-merger part of neutron-star mergers. In this case, the waveform inferred would provide us with a direct way to look for deconfinement to quark matter \cite{Most:2018eaw}. But, in order to know what to look for in such signals, and how to compare the results with what is already known from heavy-ion collisions (see Refs.~\cite{MUSES:2023hyz,Sorensen:2023zkk,Almaalol:2022xwv,Lovato:2022vgq} for recent reviews), we need to have a better understanding on how both temperature and magnetic fields affect deconfinement, including model dependencies in the location and strength of the phase transition.

\section*{Acknowledgements}
P.~C. acknowledges support from project CERN/FIS-PAR/0040/2019. R.N.\ acknowledges financial support from CAPES, CNPq, and FAPERJ. This work is part of the project INCT-FNA Proc. No. 464898/2014-5 as well as FAPERJ JCNE Proc. No. E-26/203.299/2017. {V.~D. acknowledges support from {the Fulbright U.S. Scholar Program and} the National Science Foundation
under grants PHY1748621, MUSES OAC-2103680, and
NP3M PHY-2116686.} C.~P. and P. C. acknowledge support from FCT (Fundação para a Ciência e a Tecnologia, I.P, Portugal) under Projects  UIDP/\-04564/\-2020.  UIDB/\-04564/\-2020 and 2022.06460.PTDC.

\bibliographystyle{apsrev4-2}
\bibliography{references}

\label{lastpage}
\end{document}